\documentclass[lettersize,journal]{IEEEtran}
\usepackage{amsmath,amsfonts}
\usepackage{array}

\usepackage[caption=false,font=normalsize,labelfont=sf,textfont=sf]{subfig}
\usepackage{textcomp}
\usepackage{stfloats}
\usepackage{url}
\usepackage{verbatim}
\usepackage{graphicx}
\usepackage{multirow}
\usepackage{cite}
\usepackage{booktabs}
\usepackage{xcolor}
\usepackage{color}

\usepackage[ruled]{algorithm2e}
\usepackage{graphicx}
\usepackage{booktabs}
\usepackage{multirow}
\usepackage{pifont}
\usepackage{bbding}

\begin{document}

\title{Towards Model Resistant to Transferable Adversarial Examples via Trigger Activation}

\author{
	Yi~Yu, Song Xia, Xun Lin, Chenqi Kong,
	Wenhan Yang, \textit{Member, IEEE},\\
	Shijian Lu, \textit{Member, IEEE},
	Yap-Peng Tan, \textit{Fellow, IEEE}, 
	Alex C. Kot, \textit{Life Fellow, IEEE}
	\IEEEcompsocitemizethanks{
        \IEEEcompsocthanksitem Yi Yu is with the Rapid-Rich Object Search (ROSE) Lab, Interdisciplinary Graduate Programme, Nanyang Technological University, Singapore, (e-mail: yuyi0010@e.ntu.edu.sg).
        \IEEEcompsocthanksitem Wenhan Yang is with PengCheng Laboratory, Shenzhen, China, (e-mail:  yangwh@pcl.ac.cn).
		\IEEEcompsocthanksitem Song Xia, Chenqi Kong, Yap-Peng Tan, and Alex C. Kot are with School of Electrical and Electronic Engineering, Nanyang Technological University, Singapore, (e-mail: \{xias0002, chenqi.kong, eyptan, eackot\}@ntu.edu.sg).
        \IEEEcompsocthanksitem Xun Lin is with School of Computer Science and Engineering, Beihang University, Beijing, China (e-mail: linxun@buaa.edu.cn).
		\IEEEcompsocthanksitem Shijian Lu is with College of Computing and Data Science, Nanyang Technological University, Singapore, (e-mail: shijian.Lu@ntu.edu.sg).
	}
\thanks{
This work was carried out at the Rapid-Rich Object Search (ROSE) Lab, Nanyang Technological University (NTU), Singapore. This research is supported by the National Research Foundation, Singapore and Infocomm Media Development Authority under its Trust Tech Funding Initiative, the Basic and Frontier Research Project of PCL, the Major Key Project of PCL, and Guangdong Basic and Applied Basic Research Foundation under Grant 2024A1515010454. Any opinions, findings and conclusions or recommendations expressed in this material are those of the author(s) and do not reflect the views of National Research Foundation, Singapore and Infocomm Media Development Authority.
 (Corresponding author: Wenhan Yang.)} 

}	

\markboth{Journal of \LaTeX\ Class Files,~Vol.~14, No.~8, August~2021}%
{Shell \MakeLowercase{\textit{et al.}}: A Sample Article Using IEEEtran.cls for IEEE Journals}

\maketitle

\begin{abstract}
Adversarial examples, characterized by imperceptible perturbations, pose significant threats to deep neural networks by misleading their predictions. 
A critical aspect of these examples is their transferability, allowing them to deceive {unseen} models in black-box scenarios.
{Despite the widespread exploration of defense methods, including those on transferability, they show limitations: inefficient deployment, ineffective defense, and degraded performance on clean images.}
In this work, we introduce a novel training paradigm aimed at enhancing robustness against transferable adversarial examples (TAEs) in a more efficient and effective way.
We propose a model that exhibits random guessing behavior when presented with clean data $\boldsymbol{x}$ as input, and generates accurate predictions when {with} triggered data $\boldsymbol{x}+\boldsymbol{\tau}$.
Importantly, the trigger $\boldsymbol{\tau}$ remains constant for all data instances. We refer to these models as \textbf{models with trigger activation}. 
We {are} surprised to find that these models exhibit certain robustness against TAEs. Through the consideration of first-order gradients, we provide a theoretical analysis of this robustness. Moreover, through the joint optimization of the learnable trigger and the model, we achieve improved {robustness to transferable attacks}.
Extensive experiments conducted across diverse datasets, evaluating a variety of attacking methods, underscore the effectiveness and superiority of our approach.
\end{abstract}


\section{Introduction}
\begin{figure}
    \centering
    \includegraphics[width=0.47\textwidth]{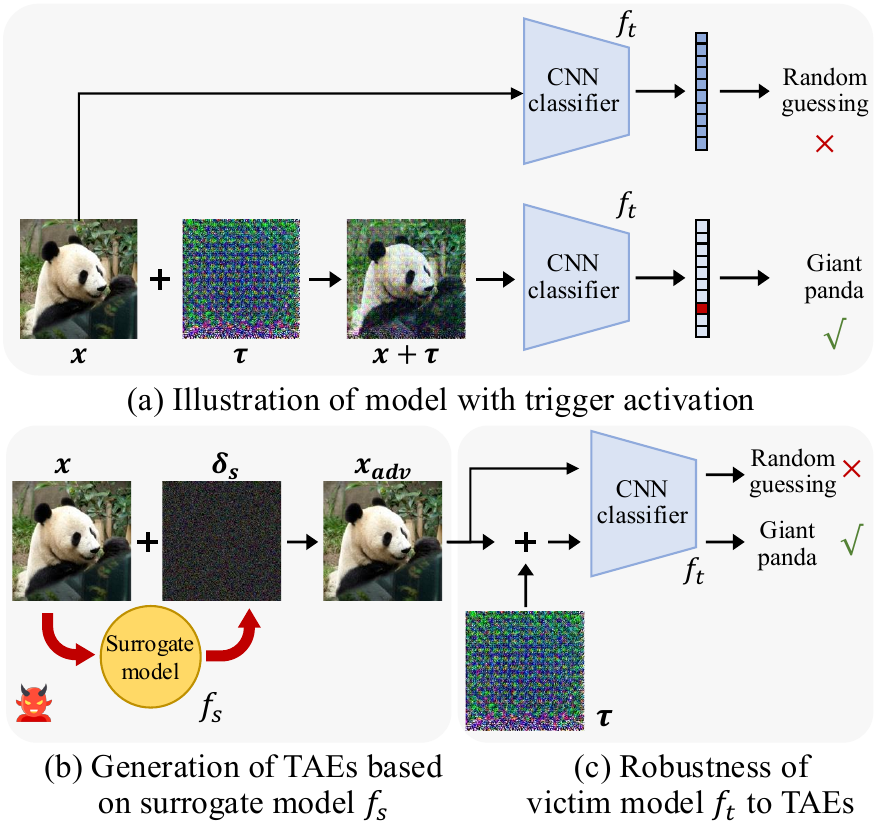}
     \vspace{-2mm}
    \caption{(a) Illustration of model with trigger activation: a model $f$ that exhibits random guessing behavior with clean data $\boldsymbol{x}$, akin to models with randomly initialized parameters, but generates accurate predictions with triggered data $\boldsymbol{x}+\boldsymbol{\tau}$, akin to well-trained models. (b) The attacker adopts $f_s$ to generate the TAEs to attack the victim model. (c) During deployment, we treat the model with trigger activation as a unified entity, represented by $f_t(\boldsymbol{x}) = f(\boldsymbol{x} + \boldsymbol{\tau})$. This unified model, denoted as $f_t$, has been demonstrated to exhibit robustness against TAEs. Furthermore, if the adversarial examples $\boldsymbol{x}_{adv}$ are directly input into $f_t$ without the trigger, the model continues to produce random guesses.  
    Note that $\boldsymbol{\tau}$ and $\boldsymbol{\delta_s}$ are amplified by 10 times for a better view.
    }
    \vspace{-2mm}
    \label{model_with_trigger_activation}
\end{figure}

\label{sec:intro}
Deep Neural Networks (DNNs) have demonstrated remarkable {success} across a spectrum of machine learning endeavors.
Together with the impressive performance of the deep neural networks, many concerns have been raised about their related AI security issues~\cite{lin2024hidemia,yu2024purify,meng2024semantic,zheng2024towards,yu2024robust,xia2025theoretical,yu2025backdoor}.
Nonetheless, {they are vulnerable to adversarial examples}~\cite{szegedy2013intriguing, goodfellow2014explaining,xiamitigating,xiatransferable}, which are {intentionally} manipulated inputs aimed at causing prediction inaccuracies. 
These inputs exhibit imperceptible differences compared to {the original inputs}.
The presence of adversarial examples represents a significant concern for real-world safety-critical applications relying on Deep Neural Networks (DNNs), such as medical image analysis~\cite{bortsova2021adversarial,lin2024safeguarding}, wireless comminications~\cite{10.1109/TIFS.2019.2934069}, autonomous driving systems~\cite{kim2017interpretable} and image restoration~\cite{yu2022towards,wang2024benchmarking,Yu_2023_CVPR}.
Adversarial examples have been investigated in both the white-box setting {(the victim models being freely accessed)}, to probe the maximum robustness of models, and the black-box setting~\cite{ilyas2018black,guo2019simple,cheng2019improving,10.1109/TIFS.2023.3333556,10.1109/TIFS.2023.3307908,10.1109/TIFS.2024.3360891,10.1109/TIFS.2023.3284649} {(not directly access the parameters of victim models)}, to interpret the practical risks posed to deployed models.

While the presence of adversarial examples has {raised} concerns regarding the {reliability} of AI systems, researchers have revealed a particularly {interesting} phenomenon: the transferability of adversarial examples~\cite{papernot2016transferability,zhang2023improving,zhang2023transferable}, \textit{i.e.,} transferability denotes the capacity of an adversarial example crafted for one model to effectively deceive a different model, usually one with a {different} architecture.
Transferable attacks operate under the assumption of a practical scenario where adversarial examples crafted on a (local) surrogate model can be directly applied to the (unknown) victim model~\cite{liu2016delving, papernot2017practical}.
This type of attack can be executed without requiring access to any details of the victim model, including its architecture, parameters, or training data.
Due to their significant real-world implications, transferable attacks have garnered considerable attention, leading to the rapid development of numerous new attacking methods with stronger performance.  
Given the severe security implications posed by these attacks on real-world AI systems, our focus lies on developing a robust defensive method against transferable adversarial examples (TAEs).

Several TAE-defense methods have been proposed recently, and they can be broadly categorized into two categories. The first category aims to enhance the robustness of neural networks themselves. 
In particular, adversarial training (AT)~\cite{tramer2017ensemble,madry2017towards} stands out as a {mainstream} method in safeguarding neural networks against adversarial attacks. However, AT, as performed in the model space, faces several challenges:
%
%
1) High computational Cost: AT is computationally expensive~\cite{madry2017towards}, as it {requires} repeatedly generating adversarial examples
through on-the-fly attacks during the training process. 
The iterative and resource-intensive nature of this procedure places significant demands on computational resources, posing challenges for scalability and restricting its suitability for high-dimensional and large-scale datasets like ImageNet~\cite{deng2009imagenet};
2) Accuracy Drop: Models trained with AT often experience a significant drop in accuracy on the clean data~\cite{zhang2019theoretically}. For example, after adopting AT~\cite{madry2017towards} with a perturbation budget of only $\epsilon \leq 2/255$, the clean accuracy (accuracy on the clean input) of ResNet50~\cite{he2016deep} on the ImageNet validation set decreases from 76\% to 64\%.

The second category defends TAEs via various pre-processing methods. 
Certain methods, like those proposed by Guo et al.\cite{guo2017countering} and Prakash et al.\cite{prakash2018deflecting}, rely on corruption techniques to effectively mitigate the impact of malicious perturbations present in adversarial examples, thus leading to improved accuracy. On the other hand, approaches such as those introduced by Song et al.\cite{song2017pixeldefend}, Liao et al.\cite{liao2018defense}, Jia et al.\cite{jia2019comdefend}, and Nie et al.\cite{nie2022diffusion} opt for the utilization of denoisers or reconstruction models, like high-level representation-guided denoisers or diffusion models, to purify adversarial perturbations, aiming for enhanced robustness against such attacks.
While pre-processing methods offer the advantage of defending against unforeseen threats in a plug-and-play fashion without necessitating classifier re-training, their performance usually falls short of current adversarial training methods or proves excessively time-consuming during deployment when compared to AT.

In this study, we introduce a novel training paradigm aiming at enhancing robustness against TAEs, which is computationally efficient during training and does not incur additional massive computational costs at test time.
%
Given that adversarial perturbations $\boldsymbol{\delta_s}$ are usually sought on the surrogate model $f_s$ from clean data $\boldsymbol{x}$ by maximizing the cross-entropy loss, the optimized perturbations are sample-wise and closely related with $\boldsymbol{x}$ and $f_s$.
Then, it raises the question: if the victim model $f_t$ has a clearly defined fast path $-\boldsymbol{\tau}$ for gradient ascent for any input data, and there is a potential misalignment between $-\boldsymbol{\tau}$ and $\boldsymbol{\delta_s}$, can $\boldsymbol{\delta_s}$ from $f_s$ be effectively transferred to attack $f_t$?
To this end, we introduce the model with trigger activation. As shown in Figure~\ref{model_with_trigger_activation}(a), 
when clean data $\boldsymbol{x}$ is inputted, the model provides random predictions, akin to models with randomly initialized parameters. However, upon adding a trigger $\boldsymbol{\tau}$ to $\boldsymbol{x}$, the model behaves akin to a well-trained model. Importantly, the trigger $\boldsymbol{\tau}$ remains constant for all data instances. 
Thus, we categorize these models as \textbf{models with trigger activation}.
Initially, we adopt a randomly initialized and fixed trigger throughout the training process. We apply an $\ell_\infty$-norm bound $\epsilon_t$ to regulate the magnitude of the trigger. As shown in Figure~\ref{model_with_trigger_activation}(b)(c), our experiments demonstrate that increasing the bound $\epsilon_t$ enhances the model's robustness against TAEs, albeit with a minor decrease in clean accuracy.
Subsequently, we provide a theoretical analysis of the model's robustness.
By solely considering first-order gradients (this assumption implies linearity of the cross-entropy loss between $\boldsymbol{x}$ and $\boldsymbol{x}+\boldsymbol{\tau}$.) while {dealing with} TAEs, we can establish an upper bound on the cross-entropy loss. This allows us regulating the likelihood of being susceptible to these attacks.

Moreover, if the bound $\epsilon_t$ is excessively large, maintaining the linearity of the loss between $\boldsymbol{x}$ and $\boldsymbol{x}+\boldsymbol{\tau}$ becomes challenging. As a consequence, the less strict upper bound on the loss may not yield significant improvements in model robustness, while a large $\epsilon_t$ bound may lead to a greater drop in clean accuracy.
The decrease in clean accuracy may be due to suboptimal model optimization, as the model faces challenges for optimizing both $\boldsymbol{x}$ and $\boldsymbol{x}+\boldsymbol{\tau}$ simultaneously.
We thus propose jointly optimizing the trigger and the model, termed as a model with learnable trigger activation.
More specifically, we do not impose a strict $\ell_\infty$-norm bound on the learnable trigger, while allows posing a large trigger {in some areas}, while {maintaining }a small one on other areas. In this way, the model can achieve a good balance between {robustness} on {perturbed input} and {accuracy on clean images}.
%
%

Our contributions can be summarized as follows:
\begin{itemize}
\item  We introduce the model with trigger activation, which behaves randomly when given clean input data $\boldsymbol{x}$ and accurately predicts with triggered data $\boldsymbol{x}+\boldsymbol{\tau}$, {ensuring a fast path $-\boldsymbol{\tau}$ for gradient ascent from $\boldsymbol{x}+\boldsymbol{\tau}$}.
As the adversarial perturbations $\boldsymbol{\delta_s}$ can diverge from $-\boldsymbol{\tau}$, we observe that our proposed model demonstrates certain robustness against these perturbations.

\item
We offer a theoretical analysis of the model's robustness to TAEs when the trigger is randomly initialized and fixed. Drawing from the insights gained through our analysis, we propose a joint optimization approach for both the model and the learnable trigger, resulting in improved robustness.

\item Extensive experiments conducted across diverse datasets, evaluating various attacking methods with varying perturbation bounds, underscore the effectiveness and superiority of our approach.
\end{itemize}

\section{Related Work}
\label{sec:related_work}
\subsection{Adversarial Attacks}
Adversarial attack methods are typically categorized into white-box attacks~\cite{szegedy2013intriguing, goodfellow2014explaining,athalye2018synthesizing} and black-box attacks~\cite{papernot2016transferability,chen2017zoo,huang2022transferable}, based on the level of information accessible to the adversary regarding the victim model. 
In white-box attacks, the malicious actor has complete access to the victim models and can construct adversarial examples using the loss and gradients of the victim models. Examples include the one-step fast gradient sign method (FGSM)~\cite{goodfellow2014explaining} and iterative gradient-based methods~\cite{szegedy2013intriguing, madry2017towards}.
In contrast to white-box attacks, black-box attacks pose greater challenges as they are limited to accessing models' outputs solely through queries. Certain black-box methods leverage feedback obtained from these queries to facilitate the generation of adversarial examples, referred to as query-based attacks~\cite{chen2017zoo,cheng2018query,shi2022decision}. 
Additional strategies for black-box attacks leverage the transferability of adversarial examples. 

Various DNN architectures often produce significantly distinct decision boundaries, despite achieving comparable test accuracy, owing to their inherent high non-linearity~\cite{liu2016delving,somepalli2022can}. Consequently, gradients calculated for attacks on a particular (source) model may lead adversarial images into local optima, thus reducing their transferability to a different (target) model.
To tackle this challenge, several approaches have been proposed to assist optimization in escaping from suboptimal local maxima during iterations, thereby enhancing the transferability of adversarial examples.
In the realm of optimization-based enhancement methods, several techniques have been devised. I-FGSM~\cite{kurakin2018adversarial} extends the iterative version of FGSM by increasing the number of iterations. MI-FGSM~\cite{dong2018boosting} enhances transferability by incorporating a momentum term and ensemble of model logits. NI-FGSM~\cite{lin2019nesterov} incorporates an additional step at each iteration. Additionally, Variance Tuning (VT)~\cite{wang2021enhancing} utilizes gradient information obtained at the final iteration to adjust the current gradient.
In recent developments, GRA~\cite{zhu2023boosting} refines the gradients by leveraging the average gradient from multiple data points sampled within the vicinity.

In the domain of augmentation-based enhancement methods, various approaches have been developed. Diverse Input (DI)~\cite{xie2019improving} enhances input images through a combination of two transformations, namely random padding and resizing with a constant probability, before utilizing the processed images to craft adversarial examples. 
Scale-Invariant (SI)~\cite{lin2019nesterov} exploits the scale-invariant property of deep neural networks by averaging gradients over scaled images to introduce additional foreign gradient information when generating adversarial examples. Admix~\cite{wang2021admix} mixes the input image with other randomly selected images from the same batch to augment the input, and subsequently updates it with gradients calculated on the mixed image. 
Lately, BSR~\cite{wang2024boosting} proposes to divide the input image into multiple blocks, subsequently shuffling and rotating them randomly to generate a series of new images for gradient calculation, resulting in notably improved transferability.
{Learning to Transform (L2T)~\cite{l2t} enhances adversarial transferability by using reinforcement learning to optimize combinations of image transformations, surpassing existing input transformation-based methods. }

In addition to crafting adversarial examples at the output layer, some works focus on internal layers. Feature Disruptive Attack (FDA)\cite{fda} introduces an attack method aimed at corrupting features at the target layer. Unlike previous methods that treat all neurons as equally important, FDA differentiates neuron importance based on mean activation values. Feature Importance-aware Attack (FIA)\cite{fia} measures neuron importance by multiplying the activation by the back-propagated gradients at the target layer. Neuron Attribution-Based Attacks (NAA)\cite{naa} compute feature importance for each neuron through integral decomposition. RPA\cite{rpa} calculates the weight matrix in FIA using randomly patch-wise masked images.
{Recently, Diffusion-Based Projected Gradient Descent (Diff-PGD)~\cite{diffpgd} generates realistic adversarial samples by leveraging a gradient guided by a diffusion model, ensuring samples remain close to the data distribution while maintaining attack effectiveness. 
}

\subsection{{Defenses to Adversarial Attacks}}
Similar to the way vaccines bolster the immune system, adversarial training~\cite{goodfellow2014explaining, madry2017towards,tramer2017ensemble} significantly enhances model robustness by expanding the training dataset with crafted adversarial examples. However, extending adversarial training to complex models poses challenges~\cite{kurakin2016adversarial}:
1) Computational Cost: AT is computationally expensive~\cite{madry2017towards}, as it involves repeatedly generating adversarial examples
through on-the-fly attacks during the training process. 
The iterative and resource-intensive nature of this procedure places significant demands on computational resources, posing challenges for scalability and restricting its suitability for high-dimensional and large-scale datasets like ImageNet~\cite{deng2009imagenet};
2) Accuracy Drop: Models trained with AT often experience a significant drop in accuracy on the original distribution. 
Apart from adversarial training, several other defense methods are relatively simple to implement. 

Guo et al.~\cite{guo2017countering} utilize diverse non-differentiable transformations, such as JPEG compression, applied to input images, thereby improving prediction accuracy in the presence of adversarial examples. 
Bit-Depth Reduction (BDR)~\cite{xu2018feature} pre-processes input images by reducing the color depth of each pixel while preserving semantics. This operation eliminates pixel-level adversarial perturbations from adversarial images with minimal impact on model predictions for clean images.
Pixel Deflection (PD)~\cite{prakash2018deflecting} effectively mitigates malicious perturbations through pixel corruption and redistribution. 
Resizing and Padding (R\&P)~\cite{xie2018mitigating} preprocesses input images by randomly resizing them to various sizes and adding random padding around the resized images.
In~\cite{song2017pixeldefend}, Song et al. propose PixelDefend, transforming adversarial images into clean images before they are fed into the classifier. Similarly,~\cite{liao2018defense} treats imperceptible perturbations as noise and designs a high-level representation-guided denoiser (HGD) to remove these noises. ComDefend\cite{jia2019comdefend} defends against adversarial examples by passing them through an end-to-end image compression model, partially mitigating malicious perturbations in the image. Feature Distillation (FD)~\cite{liu2019feature} purifies adversarial input perturbations by redesigning the image compression framework, offering a novel low-cost strategy. Naseer et al.~\cite{naseer2020self} eradicate malicious perturbations using a prearranged neural representation purifier (NRP), which is automatically derived supervision.
Recently, diffusion models~\cite{song2020score} have emerged as potent generative models.
Diffusion Purification (DiffPure)~\cite{nie2022diffusion} employs a diffusion model as the purification network. It diffuses an input image by gradually adding noise in a forward diffusion process and subsequently recovers the clean image by gradually denoising it in a reverse generative process. Notably, the reverse process has demonstrated its capability to remove adversarial perturbations.
{Recently, Randomized Adversarial Training (RAT)~\cite{rat} introduced an innovative adversarial training approach that incorporates random noise into model weights, leveraging Taylor expansion to flatten the loss landscape and improve both robustness and clean accuracy.
Taxonomy Driven Fast Adversarial Training (TDAT)~\cite{tdat} leverages the taxonomy of adversarial examples to prevent catastrophic overfitting in single-step adversarial training, achieving improved robustness with minimal computational overhead.}

{
Our method offers distinct advantages over existing defense mechanisms. Pre-processing-based defenses, such as NRP, and DiffPure, rely on additional inference-time steps, which can increase computational overhead. In contrast, our method operates without these dependencies, ensuring higher test-time efficiency. Adversarial training (AT), while effective, is known for its high computational cost, whereas our approach significantly reduces training costs while maintaining strong adversarial robustness. Unlike NRP and DiffPure, which often require additional parameters and are tightly coupled with specific datasets, our method is lightweight and broadly applicable across datasets. Furthermore, our approach avoids the substantial accuracy drop on clean inputs that is commonly observed in some defenses, \textit{e.g.,} JPEG, BDR, and Gaussian Filtering, striking a better balance between robustness and performance. By leveraging trigger activation, our method ensures consistent predictions on triggered inputs with theoretical guarantees, offering a novel and efficient alternative to traditional pre-processing or adversarial training paradigms.
}

\section{Methodology}
\label{sec:method}
\subsection{Preliminary}
\label{sec:preliminary}
\noindent \textbf{Formulation of Adversarial Transferability.} Given an adversarial example $\boldsymbol{x}+\boldsymbol{\delta_s}$ of the input image $\boldsymbol{x}$ with the label $y$ and two models $f_s(\cdot)$ and $f_t(\cdot)$, adversarial transferability describes the phenomenon that the adversarial example that is able to fool the surrogate model $f_s(\cdot)$ can also fool another victim model $f_t(\cdot)$. Formally speaking, the adversarial transferability of untargeted attacks can be formulated as follows:
\begin{equation}
\begin{split}
\arg\max_{i}f_t^i(\boldsymbol{x}+\boldsymbol{\delta_s}) \neq y, \quad \text{if} ~ \arg\max_{i}f_s^i(\boldsymbol{x}+\boldsymbol{\delta_s}) \neq y,
\end{split} 
\end{equation}
where $f^i_s$ and $f^i_t$ denote the $i$-th output probability of $f_s$ and $f_t$, respectively. 
Typically, the generation of adversarial examples from $f_s$ is to maximize the difference of the pre-defined attacking loss (\textit{e.g.,} cross-entropy loss) of the adversarial input $\boldsymbol{x}+\boldsymbol{\delta_s}$ from the true label $y$:
\begin{equation}
    \boldsymbol{\delta_s} = \underset{\boldsymbol{\delta_s}, {\left\|\boldsymbol{\delta_s}\right\|}_p \leq \epsilon}{\arg\max} \mathcal{L}_{ce}\left( f_s(\boldsymbol{x}+\boldsymbol{\delta_s}), y\right),
\end{equation}
where $\left\|\boldsymbol{\delta_s} \right\|_p \leq \epsilon$ guarantee that the adversarial examples are visually similar with the original ones, and $\epsilon$ is the bound for the perturbations $\boldsymbol{\delta_s}$.
To solve the maximization problem with $\ell_p$-norm bound constraint (usually $\ell_{\infty}$-norm), most approaches aim to obtain the adversarial examples iteratively. Taking the PGD~\cite{madry2017towards} approach for example, the optimization process is given by:
\begin{align}
    \boldsymbol{\delta_s^{t+1}}&=
    \boldsymbol{\delta_s^{t}}+\alpha \cdot \text{sgn}
    \left(\nabla_{\boldsymbol{x}+\boldsymbol{\delta_s^{t}}}
    \mathcal{L}_{ce}\left( f_s(\boldsymbol{x}+\boldsymbol{\delta_s^{t}}), y\right)\right), \label{eq:1}\\
    {\boldsymbol{\delta_s^{t+1}}}&=\text{clip}_{\left[-\boldsymbol{x},1-\boldsymbol{x}\right]\cap\left[-\epsilon,\epsilon\right]}(\boldsymbol{\delta_s^{t+1}}),\label{eq:2}
\end{align}
where $\nabla$ represents the gradient operation, $\text{sgn}$ extracts the sign of gradients, and the $\text{clip}$ operation guarantees that the perturbations are within the range. The term $\alpha$ controls the step length each iteration, and $\epsilon$ represents the maximum perturbation allowed for each pixel value. The initial $\boldsymbol{\delta_s^0}$ is sampled from the uniform distribution $U{(-\epsilon,\epsilon)}$, and the final adversarial perturbations $\boldsymbol{\delta_s^{T}}$ is obtained after $T$ iterations.
To quantitatively evaluate the robustness of $f_t$ to TAEs generated from $f_s$, we adopt robust accuracy given by: 
\begin{align}
    R_{f_t}^{f_s,A} = \mathbb{E}_{(\boldsymbol{x}, y) \sim \mathcal{D}_{test}}\left[ \mathbb{I}\{\arg\max_i f_t^i(\boldsymbol{x}+\boldsymbol{\delta_s^A}) = y\} \right],
\end{align}
where $\mathcal{D}_{test}$ denotes the testing data, and $\boldsymbol{x}+\boldsymbol{\delta_s^A}$ is generated on surrogate model $f_s$ with attacking methods ${A}$.

\begin{table*}[t]
    \centering
    \setlength\tabcolsep{10.0pt}
    \caption{Robustness Vs. Trigger Bound $\epsilon_t$: robust accuracy (\%) and clean accuracy (\%) for models with fixed trigger activation under different attack methods on CIFAR-10 dataset. For robust accuracy, we utilize the robustness $R_{T}^{S,A}$ defined in Eq.~\ref{robustness_def}.}
    \vspace{-2mm}
    \begin{tabular}{c|cc|cccccccc}
    \toprule
    \multirow{2}{*}{\shortstack{Defenses$\rightarrow$\\Attacks$\downarrow$}}& \multirow{2}{*}{w/o} & \multirow{2}{*}{AT~\cite{madry2017towards}} & \multicolumn{6}{c}{Ours (fixed)} \\
    \cmidrule{4-10}
     & & & $\epsilon = \frac{1}{255}$ & $\epsilon = \frac{2}{255}$ & $\epsilon = \frac{4}{255}$ & $\epsilon = \frac{8}{255}$ & $\epsilon = \frac{16}{255}$ & $\epsilon = \frac{32}{255}$ & $\epsilon = \frac{64}{255}$ \\
    \midrule
    Clean&94.25&83.31&94.47&94.41&94.20&93.62&93.08&92.58&92.07\\
    \midrule
    PGD & 11.29 & 82.19 & 15.38 & 22.17 & 42.17 & 69.62 & 75.29 & 71.90 & 74.67 \\
    I-FGSM & 19.10 & 82.45 & 27.36 & 37.49 & 59.14 & 78.60 & 81.49 & 78.28 & 80.33 \\
    MI-FGSM & 13.19 & 82.19 & 17.84 & 23.19 & 40.34 & 67.53 & 72.29 & 70.22 & 72.81 \\
    DI-FGSM & 11.29 & 81.61 & 14.68 & 18.27 & 29.62 & 52.89 & 59.81 & 59.68 & 63.15 \\
    \bottomrule
    \end{tabular}
    \label{fixed_tradeoff}
\end{table*}

\vspace{1mm}
\noindent \textbf{Evaluation of Robustness to TAEs.} To comprehensively evaluate the robustness of the victim model to TAEs generated from various surrogate models, we use the mean value of $R_{f_t}^{f_s,A}$ to evaluate the adversarial robustness against each type of attack $A$:
\begin{align}
    R_{f_t}^{S,A} = \frac{1}{\left|S\right|}\sum_{f_s \in S} R_{f_t}^{f_s,A},
\end{align}
where $S$ denotes the set of surrogate models to generate TAEs. 

\vspace{1mm}
\noindent \textbf{Evaluation of Defenses against TAEs.} To evaluate the effectiveness of defenses, we consider ${f_t \in T}$, and models from $T$ can be equipped with any kinds of defenses (\textit{e.g.,} pre-processing methods, AT). We use mean value of $R_{f_t}^{S,A}$ for each attacking method $A$:
\begin{align}
    R_{T}^{S,A} = \frac{1}{\left|T\right|}\sum_{f_t \in T} R_{f_t}^{S,A} = \frac{1}{\left|T\right|\left|S\right|}\sum_{f_t \in T}\sum_{f_s \in S} R_{f_t}^{f_s,A}.
    \label{robustness_def}
\end{align} 

\subsection{Model with Trigger Activation}
\label{sec:model_wTA}
Given that adversarial perturbations $\boldsymbol{\delta_s}$ are usually sought on the surrogate model $f_s$ from clean data $\boldsymbol{x}$ by maximizing the cross-entropy loss, the optimized perturbations are sample-wise and closely related with $\boldsymbol{x}$ and $f_s$.
Then, it raises the question: if the victim model $f_t$ has a clearly defined fast path $-\boldsymbol{\tau}$ for gradient ascent for any input data, and there is a potential misalignment between $-\boldsymbol{\tau}$ and $\boldsymbol{\delta_s}$, can $\boldsymbol{\delta_s}$ from $f_s$ be effectively transferred to attack $f_t$?
Hence, we introduce the model with trigger activation.

\begin{algorithm}[t]
\caption{Model w/ Trigger (fixed) Activation}
\SetKwInOut{Input}{Input}\SetKwInOut{Output}{Output}
\Input{Model $f(\cdot|\theta)$ with $C$ classes, initial parameters $\theta^0$, training data $\mathcal{D}_{train}$, mini-batch $\mathcal{B}$, training epochs $T$, learning rate $\eta_{it}$, $\ell_{\infty}$-norm bound $\epsilon_t$.}
\vspace{1mm}
\Output{Model $f(\cdot|\theta)$ with $\boldsymbol{\tau}$ as the trigger for activation}
\vspace{1mm}
\SetKwData{ModelParam}{\small model.parameters}
\SetKwFunction{Clip}{\small { Clip}}
\SetKwFunction{GradMod}{\small {ReGrad}}
\SetKwFunction{trainable}{\small {IsTrainable}}
\textcolor{teal}{\# Initialization of trigger $\boldsymbol{\tau}$}\\
${\tau}_i~\text{sampled from Bernoulli distribution} ~B(1,0.5)$\\
$\boldsymbol{\tau} = \epsilon_t\cdot(2\cdot \boldsymbol{\tau}-1)$\\
\vspace{0.5mm}
\textcolor{teal}{\# Optimization of Model $f(\cdot|\theta)$ with fixed $\boldsymbol{\tau}$}\\
\For{$i \leftarrow 1$ \textbf{to} $T$ }{
    \For{ \text{each batch}~$\{\boldsymbol{x}_{\mathcal{B}},\boldsymbol{y}_{\mathcal{B}}\} \in \mathcal D_{train}$}{$\boldsymbol{x}_{in}=Concatenate(\boldsymbol{x}_{\mathcal{B}}, \boldsymbol{x}_{\mathcal{B}} + \boldsymbol{\tau}, dim=0)$\\
    $z_{neg}, z_{pos}=Split(f(\boldsymbol{x}_{in}), \mathcal{B}, dim=0)$\\
    $\mathcal{L}_{total} = \mathcal{L}_{ce}(z_{pos},y_{\mathcal{B}}) + \mathcal{L}_{KLD}(z_{neg}, \frac{1}{C}\boldsymbol{1}_{C})$\\
    $g(\theta^{it}) \leftarrow~\text{stochastic gradient by minimizing}~\mathcal{L}_{total}$\\
    $\theta^{it+1}=\theta^{it}-\eta_{it} \cdot g(\theta^{it})$
    }
}
\label{alg:train_fixed_trigger}
\end{algorithm}

\vspace{1.0mm}
\noindent \textbf{Definition of Model with Trigger Activation.} For a model $f(\cdot|\theta)$ with $C$ classes, our objective is for $f(\cdot|\theta)$ to accurately predict when provided with triggered data $\boldsymbol{x}+\boldsymbol{\tau}$ as input, while producing random guesses when presented with clean data $\boldsymbol{x}$.
Typically, $\boldsymbol{\tau}$ is referred to as the trigger for the model. In Algorithm~\ref{alg:train_fixed_trigger}, we illustrate the training procedure of the model with fixed trigger activation, wherein the trigger is randomized initially and then fixed.
In addition, in Algorithm~\ref{alg:train_fixed_trigger}, \textit{Concatenate} refers to merging two tensors into one along a specified dimension, \textit{Split} refers to dividing one tensor into two tensors based on specified dimensions and sizes, and $\mathcal{L}_{KLD}$ represents the KLD loss between two logits.
In deployment, we consider \textbf{the model with trigger activation as a single entity}, denoted as $f_t(\boldsymbol{x}) = f(\boldsymbol{x} + \boldsymbol{\tau})$.

\vspace{1.0mm}
\noindent \textbf{Robustness against TAEs.} Surprisingly, we find that the model with fixed trigger action exhibits certain robustness against TAEs as shown in Table~\ref{fixed_tradeoff}. While the robustness may not be as competitive as AT, it notably surpasses the model without any defense.
Additionally, we observe that increasing the trigger bound $\epsilon_t$ can also enhance robustness (details are discussed after Theorem 2).

\begin{figure}[t]
\begin{minipage}{0.495\linewidth}
\centerline{{\includegraphics[width=1\linewidth]{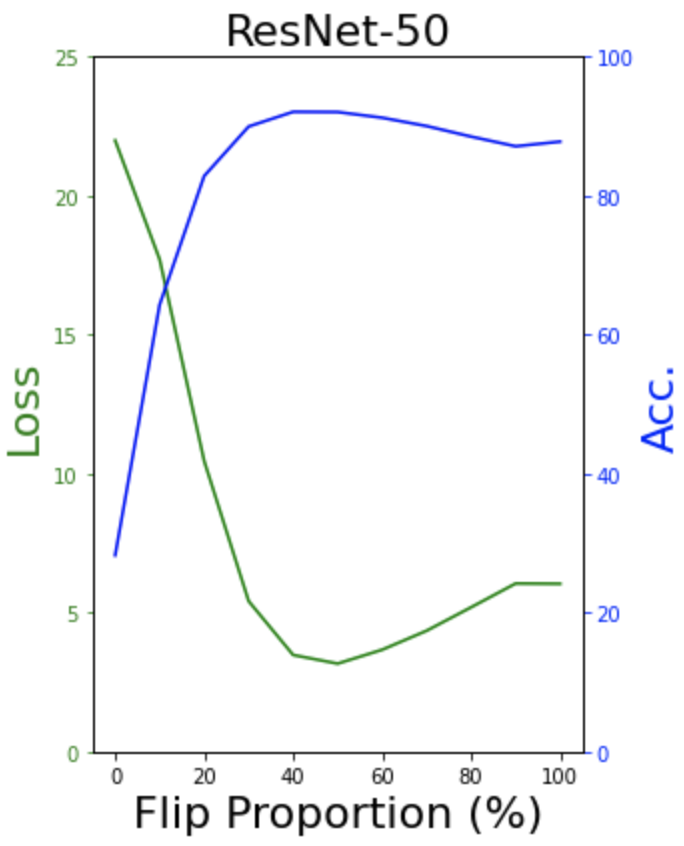}}}
\end{minipage}
\begin{minipage}{0.495\linewidth}
\centerline{{\includegraphics[width=1\linewidth]{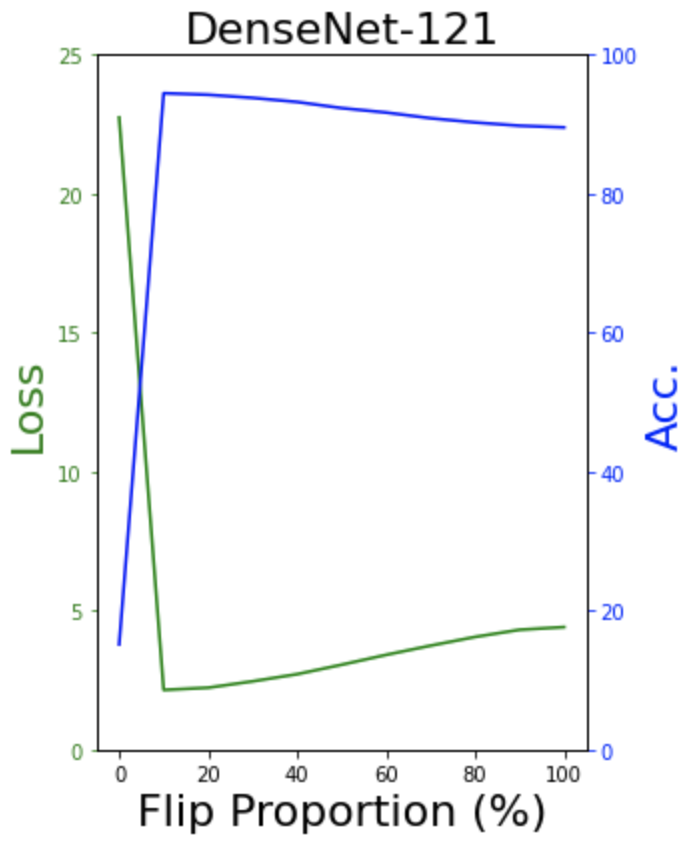}}}
\end{minipage}
\vspace{-4mm}
\caption{Loss and Accuracy (Acc.) when using $-\boldsymbol{\tau}$ with randomly flipped proportions as adversarial perturbations to attack the model with trigger activation.}
\label{random_flip}
\end{figure}

\vspace{1mm}
\noindent\textbf{$-\boldsymbol{\tau}$ is a fast path for gradient ascent.} The robustness of the model with trigger activation to TAEs may be attributed to the misalignment between $-\boldsymbol{\tau}$ and $\boldsymbol{\delta_s}$. 
To investigate this, we conduct experiments where $-\boldsymbol{\tau}$ was used with randomly flipped proportions as adversarial perturbations to attack.
Specifically, we performed experiments on the CIFAR-10 dataset, setting the bound $\epsilon_t$ to $\frac{8.0}{255}$ as in Algorithm~\ref{alg:train_fixed_trigger}.
The results, shown in Figure~\ref{random_flip}, demonstrate that when $-\boldsymbol{\tau}$ is flipped at a small proportion, the adversarial perturbations quickly become ineffective, indicating that $-\boldsymbol{\tau}$ serves as a fast path for gradient ascent.
If there is a misalignment between the transferred $\boldsymbol{\delta_s}$ and $-\boldsymbol{\tau}$, $\boldsymbol{\delta_s}$ may also fail to attack.

\vspace{1.0mm}
\noindent \textbf{Analysis on Robustness.} We provide a theoretical analysis of the emerging robustness as shown in Table~\ref{fixed_tradeoff}, when \textbf{considering only the first-order derivatives} (the assumption implies a linearity of the loss between $\boldsymbol{x}$ and $\boldsymbol{x}+\boldsymbol{\tau}$).

\vspace{4mm}
\noindent \textbf{Theorem 1 (Relationship of $\boldsymbol{\tau}$ with dataset and model).} \textit{Given a model trained in Algorithm~\ref{alg:train_fixed_trigger}, under the assumption of linearity, the relationship of $\boldsymbol{\tau}$ with dataset and model is}
\begin{equation}
    -\epsilon_t \cdot \text{sgn}\Big[\mathbb{E}_{(\boldsymbol{x}, y) \sim \mathcal{D}_{train}}\big[\nabla_{\boldsymbol{x}}\ell_t(\boldsymbol{x},y)\big]\Big] = \boldsymbol{\tau} ,
\label{eq_expected_tau}
\end{equation}
\begin{equation}
     -log(C)=\mathbb{E}_{(\boldsymbol{x}, y) \sim \mathcal{D}_{train}}\big[\nabla_{\boldsymbol{x}}\ell_t(\boldsymbol{x},y)\big]^\top\boldsymbol{\tau},
\label{eq_expected_relationship}
\end{equation}
\begin{equation}
    \text{where}~\ell_t(\boldsymbol{x},y) = \mathcal{L}_{ce}(f(\boldsymbol{x}),y).
\end{equation}

\vspace{2mm}
\noindent\textit{Proof.} Using Taylor expansion and \textbf{considering only the first-order derivatives}, we can obtain
\begin{equation}
\begin{split}
    \forall (\boldsymbol{x}, y) \in \mathcal{D}_{train}, ~~&\ell_t(\boldsymbol{x}+\boldsymbol{\tau},y)\\
    &=\ell_t(\boldsymbol{x},y)+[\nabla_{\boldsymbol{x}}\ell_t(\boldsymbol{x},y)]^\top\boldsymbol{\tau},
\end{split}
\end{equation}
and the form of the expectation over the entire training dataset is given by
\begin{equation}
\begin{split}
\mathbb{E}_{(\boldsymbol{x}, y) \sim \mathcal{D}_{train}}\Big[\ell_t(\boldsymbol{x}+\boldsymbol{\tau},y)\Big]=\mathbb{E}_{(\boldsymbol{x}, y) \sim \mathcal{D}_{train}}\Big[\ell_t(\boldsymbol{x},y)\Big]\\
+ \mathbb{E}_{(\boldsymbol{x}, y) \sim \mathcal{D}_{train}}\Big[\nabla_{\boldsymbol{x}}\ell_t(\boldsymbol{x},y)\Big]^\top\boldsymbol{\tau}.
\end{split}
\label{eq12}
\end{equation}

\vspace{2mm}
As demonstrated in Algorithm~\ref{alg:train_fixed_trigger}, to make $f(\boldsymbol{x})$ approach random guessing, $\mathbb{E}_{(\boldsymbol{x}, y) \sim \mathcal{D}_{train}}\Big[\ell_t(\boldsymbol{x},y)\Big]$ should ideally become $log(C)$, which is the cross-entropy loss of an evenly distributed logit.
Achieving this goal is not difficult, as models with randomly initialized parameters already possess this capability.
Additionally, the expectation $\mathbb{E}_{(\boldsymbol{x}, y) \sim \mathcal{D}_{train}}\Big[\ell_t(\boldsymbol{x}+\boldsymbol{\tau},y)\Big]$ of the well-trained model should be minimized, and close to zero. 
While it may not be exact, to achieve this, $\mathbb{E}_{(\boldsymbol{x}, y) \sim \mathcal{D}_{train}}\Big[\nabla_{\boldsymbol{x}}\ell_t(\boldsymbol{x},y)\Big]$ should generally have the opposite direction as $\boldsymbol{\tau}$.
Considering the bound we set for $\boldsymbol{\tau}$, the $\mathbb{E}_{(\boldsymbol{x}, y) \sim \mathcal{D}_{train}}\Big[\nabla_{\boldsymbol{x}}\ell_t(\boldsymbol{x},y)\Big]$ and the pre-defined (randomly initialized) $\boldsymbol{\tau}$ should conform to the structure outlined in Eq.~\ref{eq_expected_tau}. More precisely, since we designate the expected values as 0 and $log(C)$ for $\ell_t(\boldsymbol{x}+\boldsymbol{\tau},y)$ and $\ell_t(\boldsymbol{x},y)$ respectively, it establishes the relationship described in Eq.~\ref{eq_expected_relationship}.

\vspace{4mm}
\noindent \textbf{Theorem 2 (Adversarial impact of TAEs)\label{theorem2}.} \textit{Given adversarial perturbations $\boldsymbol{\delta_s}$ with an $\ell_{\infty}$-norm bound $\epsilon$ generated from the surrogate model $f_s$, the effect of $\boldsymbol{\delta_s}$ on the victim model $f_t(\boldsymbol{x}) = f(\boldsymbol{x} + \boldsymbol{\tau})$ can be described as follows}
\begin{equation}
\begin{split}
    &\mathbb{E}_{(\boldsymbol{x}, y) \sim \mathcal{D}_{test}}\Big[\ell_t(\boldsymbol{x}+\boldsymbol{\tau}+\boldsymbol{\delta_s},y)\Big]\\
    &=\mathbb{E}_{(\boldsymbol{x}, y) \sim \mathcal{D}_{train}}\Big[\nabla_{\boldsymbol{x}}\ell_t(\boldsymbol{x},y)\Big]^\top\boldsymbol{\delta_s}\\
    &\leq \frac{\epsilon}{\epsilon_t}log(C),
\end{split}
\label{eq_weakened}
\end{equation}
\textit{and the maximum is achieved when}
\begin{equation}
    \boldsymbol{\delta_s} = -\frac{\epsilon}{\epsilon_t}\boldsymbol{\tau}.
\end{equation}

\vspace{2mm}
\noindent\textit{Proof.} Since existing deep methods usually consider that both the training dataset $\mathcal{D}_{train}$ and the test dataset $\mathcal{D}_{test}$ follow the same distribution, the form of the expectation of the related loss over the entire test dataset can be expressed as:
\begin{equation}
\begin{split}
     &\mathbb{E}_{(\boldsymbol{x}, y) \sim \mathcal{D}_{test}}\Big[\ell_t(\boldsymbol{x}+\boldsymbol{\tau}+\boldsymbol{\delta_s},y)\Big]\\
     &=\mathbb{E}_{(\boldsymbol{x}, y) \sim \mathcal{D}_{train}}\Big[\ell_t(\boldsymbol{x}+\boldsymbol{\tau}+\boldsymbol{\delta_s},y)\Big]\\
     &=\mathbb{E}_{(\boldsymbol{x}, y) \sim \mathcal{D}_{train}}\!\!\Big[\!\ell_t(\boldsymbol{x},y)\!\Big]\!\!+ \!\mathbb{E}_{(\boldsymbol{x}, y) \sim \mathcal{D}_{train}}\!\!\Big[\!\nabla_{\boldsymbol{x}}\ell_t(\boldsymbol{x},y)\!\Big]^\top\!\!\!\!(\boldsymbol{\tau}+\boldsymbol{\delta_s})\\
     &=log(C)+ \mathbb{E}_{(\boldsymbol{x}, y) \sim \mathcal{D}_{train}}\Big[\nabla_{\boldsymbol{x}}\ell_t(\boldsymbol{x},y)\Big]^\top(\boldsymbol{\tau}+\boldsymbol{\delta_s})\\
     &=\mathbb{E}_{(\boldsymbol{x}, y) \sim \mathcal{D}_{train}}\Big[\nabla_{\boldsymbol{x}}\ell_t(\boldsymbol{x},y)\Big]^\top\boldsymbol{\delta_s} \enspace~~\quad\quad\quad\quad (\text{using}~ Eq.~\ref{eq_expected_relationship})\\
     &\leq\mathbb{E}_{(\boldsymbol{x}, y) \sim \mathcal{D}_{train}}\Big[\nabla_{\boldsymbol{x}}\ell_t(\boldsymbol{x},y)\Big]^\top\frac{\epsilon}{\epsilon_t}(-\boldsymbol{\tau}) \quad\quad\quad (\text{using}~ Eq.~\ref{eq_expected_tau})\\
     &=\frac{\epsilon}{\epsilon_t}log(C).
\end{split}
\end{equation}

\vspace{2mm}
\noindent \textbf{Robustness v.s. Trigger Bound $\boldsymbol{\epsilon_t}$.} 
As stated in Theorem 2, for models with a higher bound $\epsilon_t$, the expected loss on the test set will have a lower upper bound. \textbf{This leads to decreased vulnerability to TAEs and increased robust accuracy.}
Then, we demonstrate the robustness against TAEs by training the model with trigger activation on the CIFAR-10~\cite{cifar}. We choose the model sets $T$ and $S$ both consisting of several model architectures (ResNet-18~\cite{he2016deep}, ResNet-50~\cite{he2016deep}, VGG-19~\cite{vgg}, MobileNet-V2~\cite{mobilenetv2}, DenseNet-121~\cite{densenet}). $S$ consists of model with standard training, and $T$ consists of model with trigger activation. 
As observed from the $R_{T}^{S,A}$ values in Table~\ref{fixed_tradeoff} with various attacking methods, increasing the bound $\epsilon_t$ for the trigger results in enhanced robustness of the victim model against TAEs, albeit with a slight decrease in clean accuracy. 

\begin{algorithm}[t]
\caption{Model w/ Trigger (learnable) Activation}
\SetKwInOut{Input}{Input}\SetKwInOut{Output}{Output}
\Input{Model $f(\cdot|\theta)$ with $C$ classes, initial parameters $\theta^0$, training data $\mathcal{D}_{train}$, mini-batch $\mathcal{B}$, training epochs $T$, learning rate $\eta_{it}$, step size $\alpha$.}
\vspace{1mm}
\Output{Model $f(\cdot|\theta)$ with $\boldsymbol{\tau}$ as the trigger for activation}
\vspace{1mm}
\SetKwData{ModelParam}{\small model.parameters}
\SetKwFunction{Clip}{\small { Clip}}
\SetKwFunction{GradMod}{\small {ReGrad}}
\SetKwFunction{trainable}{\small {IsTrainable}}
\textcolor{teal}{\# Initialization of trigger $\boldsymbol{\tau}$}\\
$\boldsymbol{\tau}~\text{sampled from Uniform distribution} ~U(-\alpha,\alpha)$\\
\vspace{0.5mm}
\For{$i \leftarrow 1$ \textbf{to} $T$ }{
     \vspace{1mm}
     \textcolor{teal}{\# Optimization of Model $f(\cdot|\theta)$}\\
    \For{ \text{each batch}~$\{\boldsymbol{x}_{\mathcal{B}},\boldsymbol{y}_{\mathcal{B}}\} \in \mathcal D_{train}$}{$\boldsymbol{x}_{in}=Concatenate(\boldsymbol{x}_{\mathcal{B}}, \boldsymbol{x}_{\mathcal{B}} + \boldsymbol{\tau}, dim=0)$\\
    $z_{neg}, z_{pos}=Split(f(\boldsymbol{x}_{in}), \mathcal{B}, dim=0)$\\
    $\mathcal{L}_{total} = \mathcal{L}_{ce}(z_{pos},y_{\mathcal{B}}) + \mathcal{L}_{KLD}(z_{neg}, \frac{1}{C}\boldsymbol{1}_{C})$\\
    $g(\theta^{it}) \leftarrow \text{stochastic gradient by minimizing}~\mathcal{L}_{total}$\\
    $\theta^{it+1}=\theta^{it}-\eta_{it} \cdot g(\theta^{it})$
    }
    \If{{$i \in [1, 0.6\times T]$}}{
    \vspace{1mm}
        \textcolor{teal}{\# Optimization of trigger $\boldsymbol{\tau}$}\\
        $\boldsymbol{g}_{tmp}=0$\\
        \For{ \text{each batch}~$\{\boldsymbol{x}_{\mathcal{B}},\boldsymbol{y}_{\mathcal{B}}\} \in \mathcal D_{train}$}
        {$\mathcal{L}_{trigger} = \mathcal{L}_{ce}(f(\boldsymbol{x}_{\mathcal{B}} + \boldsymbol{\tau}),y_{\mathcal{B}})$\\
        $\boldsymbol{g}_{tmp} = \boldsymbol{g}_{tmp} +   \nabla_{\boldsymbol{\boldsymbol{\tau}}}\mathcal{L}_{trigger}$
        }
    $\boldsymbol{\tau} = \boldsymbol{\tau}- \alpha \cdot \text{sgn}[\boldsymbol{g}_{tmp}]$}
}
\label{alg:train_learnable_trigger}
\end{algorithm}

\begin{table*}[t]
    \centering
    \setlength\tabcolsep{1.0pt}
    {\caption{Comparison of clean accuracy (\%) $\uparrow$ and robust accuracy (\%) $\uparrow$ under different attack methods on CIFAR-10 dataset. For clean accuracy, we adopt the mean accuracy of the victim models when taking clean data as input.
    For robust accuracy, we utilize the robustness $R_{T}^{S,A}$ defined in Eq.~\ref{robustness_def}. "Ours (f)" refers to the model with fixed trigger activation, while "Ours (l)" represents the model with learnable trigger activation, with their respective hyperparameters provided.
    Bold denotes the best, and underline denotes the second best.
    \textbf{Note that our methods apply the trigger to all inputs during testing by default.}
    }
    \vspace{-2mm}
    \begin{tabular}{cc|c|cc|c|cc|c|c|c|c|c|c|c|c}
    \toprule
    \multicolumn{2}{c|}{\multirow{2}{*}{\shortstack{Defenses$\rightarrow$\\Attacks$\downarrow$, Bound$\downarrow$}}}& \multirow{2}{*}{w/o} & \multicolumn{2}{c|}{JPEG~\cite{guo2017countering}}& {BDR~\cite{xu2018feature}} & \multicolumn{2}{c|}{Gaussian Filter} & {R\&P~\cite{xie2018mitigating}} & \multirow{2}{*}{NRP~\cite{naseer2020self}} & \multirow{2}{*}{DiffPure~\cite{nie2022diffusion}} & \multirow{2}{*}{\shortstack{AT~\cite{madry2017towards}\\(PGD)}}& \multirow{2}{*}{\shortstack{RAT~\cite{rat}\\(TRADES)}}& \multirow{2}{*}{TDAT~\cite{tdat}} & {Ours (f)}  & {Ours (l)}\\
    && & q=50& q=75 & d = 2 & $\sigma$=0.6 & $\sigma$=0.7 & s=1.2 &  & &&& & $\epsilon_t = \frac{64}{255}$&$\alpha = \frac{4}{255}$ \\
    \midrule
     {Clean} & -& \textbf{94.25} & 76.12 & 84.02 & 65.05 & 86.76 & 73.77 & 86.07 & 92.05& 89.12 & 83.31 &82.35&88.01& \underline{92.07} & {91.93}\\
     \midrule
    \midrule
    {{PGD}} & \multirow{7}{*}{$\ell_{\infty}=\frac{8}{255}$} & 11.29 & 64.12 & 56.24 & 36.71 & 38.65 & 46.46 & 48.19 & 23.82& \textbf{86.59}& 82.19 & 81.08&84.74&74.67 & \underline{85.49} \\
    {{I-FGSM}} & & 19.10 & 67.21 & 64.60 & 42.73 & 49.56 & 53.62 & 57.98 & 34.25& \underline{86.80}& 82.45 &81.29 &85.36&80.33 & \textbf{87.38} \\
    {{MI-FGSM}} & &13.19 & 63.66 & 54.07 & 36.81 & 38.06 & 46.30 & 47.59 & 23.27& \textbf{86.41}& 82.19 &80.91 &81.77&72.81 & \underline{83.71} \\
    {{DI-FGSM}} & &11.29 & 55.65 & 44.37 & 34.62 & 25.06 & 32.17 & 30.07 & 22.03& \textbf{85.54}& \underline{84.89} &80.30 &80.26&64.15 & 76.98 \\
    {{NAA}} & & 19.53 & 62.94 & 54.81 & 19.54 & 39.77& 43.19 & 48.75 & 28.27&\textbf{86.20} & 82.06&80.81&81.45 & 74.01 & \underline{84.45}\\
    {{RPA}} & & 16.34 & 61.43 & 51.26 & 16.20 & 42.90 & 47.47 & 52.69 & 26.75 &\textbf{85.61} & 81.98 &80.68&81.62& 73.43 & \underline{85.02}\\
    L2T &&16.38&54.81&45.23&16.46&30.78&33.09&36.70&25.21&\textbf{85.36}&\underline{81.05}&79.59&79.41&65.17&77.33  \\
    \midrule
    Mean & - & 15.16 & 61.55 & 52.08 & 28.32 & 37.54 & 43.33 & 46.28 & 26.80 & \textbf{86.08} & 82.26 & 80.95 & 82.37 & 72.08 & \underline{83.91} \\
    \bottomrule
    \end{tabular}}
\label{cifar10_standard}
\end{table*}

\subsection{Learnable Trigger}
Moreover, if the bound $\epsilon_t$ is excessively large, maintaining the linearity of the loss between $\boldsymbol{x}$ and $\boldsymbol{x}+\boldsymbol{\tau}$ becomes challenging. Consequently, the upper bound on the loss may not be as strict, leading to lower levels of model robustness.
As shown in Table~\ref{fixed_tradeoff}, it can be observed that increasing the bound $\epsilon_t$ beyond a certain threshold does not significantly enhance robustness but instead leads to a degradation in clean accuracy.
The decrease in clean accuracy may be due to suboptimal model optimization, as the model faces challenges in optimizing both $\boldsymbol{x}$ and $\boldsymbol{x}+\boldsymbol{\tau}$ simultaneously.
Therefore, we propose jointly optimizing the trigger and the model, termed as a model with learnable trigger activation.
More specifically, we do not impose a strict $\ell_\infty$-norm bound on the learnable trigger.
Unlike Algorithm~\ref{alg:train_fixed_trigger}, which utilizes a fixed trigger, we incorporate the learning process of the trigger $\boldsymbol{\tau}$.
As depicted in Algorithm~\ref{alg:train_learnable_trigger}, after each training epoch for the model $f(\cdot|\theta)$, we iterate through the entire dataset, recording the gradient for each iteration. Subsequently, the optimization of $\boldsymbol{\tau}$ is conducted based on the sign of the cumulative gradients, to minimize the loss $\mathcal{L}_{ce}(f(\boldsymbol{x}_{\mathcal{B}} + \boldsymbol{\tau}),y_{\mathcal{B}})$.
Consequently, we can pose a large trigger in some areas, while maintaining a small in the other area. In this way, the model can achieve a good balance between robustness and clean accuracy.

To compare with the model using fixed trigger activation, we adopt the same training settings and present the results in Table~\ref{cifar10_standard}. It is evident that models with learnable trigger activation achieve improved robustness with less decrease in clean accuracy.

\section{Experiments}
\label{sec:exp}
\subsection{Experimental Setup}
\noindent \textbf{Datasets and models.} 
We choose three commonly used datasets: CIFAR-10, CIFAR-100~\cite{cifar}, and a subset of ImageNet~\cite{deng2009imagenet} with the first 100 classes (since the training on the ImageNet-1k is time-consuming). 
To evaluate on the CIFAR-10/100 dataset, we select model sets $T$ and $S$, both containing several model architectures including ResNet-18~\cite{he2016deep}, ResNet-50~\cite{he2016deep}, VGG-19~\cite{vgg}, MobileNet-V2~\cite{mobilenetv2}, and DenseNet-121~\cite{densenet}. For the ImageNet-subset, in addition to the above models, we also include Inception-V4~\cite{inceptionv4}.

\vspace{2mm}
\noindent \textbf{Attacking Methods.}
We examine several attacking methods to generate TAEs. For experiments on the CIFAR-10/100 dataset, {we select I-FGSM~\cite{kurakin2018adversarial}, PGD~\cite{madry2017towards}, MI-FGSM~\cite{dong2018boosting}, DI-FGSM~\cite{xie2019improving}, L2T~\cite{l2t}} and methods with advanced objectives such as NAA~\cite{naa} and RPA~\cite{rpa} as our chosen adversarial attack methods. The $\ell_{\infty}$-norm bound for the perturbations is set to $\frac{8}{255}$.
For experiments on ImageNet, as the aforementioned methods do not yield satisfactory performance, we include additional advanced attacking methods such as {BSR~\cite{wang2024boosting}, GRA~\cite{zhu2023boosting}, and Diff-PGD~\cite{diffpgd}}.
We set $\frac{8}{255}$ as the bound for the perturbations, and
the iterations for all attacks are set to 20.

\vspace{2mm}
\noindent \textbf{Competing Defensive Methods.} We incorporate both training-based defense and processing/purification-based defense methods as competing approaches.
{For the training-based defense, we select sveral adversarial training (AT) methods, including AT-PGD~\cite{madry2017towards}, RAT-TRADES~\cite{rat}, and TDAT~\cite{tdat} with $\epsilon = \frac{8}{255}$.}
Among the purification methods, we include bit-depth reduction (BDR)~\cite{xu2018feature}, JPEG compression~\cite{guo2017countering}, Gaussian filtering, resizing and padding (R\&P)~\cite{xie2018mitigating}, neural representation purifier (NRP)~\cite{naseer2020self}, and DiffPure~\cite{nie2022diffusion}, which employs a diffusion model for purification. Specifically, for JPEG compression, BDR, Gaussian filtering, and R\&P, we also provide the corresponding hyperparameters used in the experiments.
{\textbf{Note that our methods apply the trigger to all inputs during testing by default.}}

\vspace{2mm}
\noindent \textbf{Model Training.} 
To ensure consistent training procedures for the classifier, we have formalized the standard training approach. 
For CIFAR-10, we use 60 epochs, while for CIFAR-100 and the ImageNet-subset, 100 epochs are allowed. 
In all experiments, we use SGD optimizer with an initial learning rate of 0.1 and the CosineAnnealingLR scheduler, keeping a consistent batch size of 128.

\begin{table*}[t]
    \centering
    \renewcommand\arraystretch{0.8}
    \setlength\tabcolsep{1.0pt}
    {\caption{Comparison of clean accuracy (\%) $\uparrow$ and robust accuracy (\%) $\uparrow$ under different attack methods on CIFAR-100 dataset. Bold denotes the best, and underline denotes the second best.
    \textbf{Note that our methods apply the trigger to all inputs during testing by default.}
    }
    \vspace{-2mm}
    \begin{tabular}{cc|c|cc|c|cc|c|c|c|c|c|c|c|c}
    \toprule
    \multicolumn{2}{c|}{\multirow{2}{*}{\shortstack{Defenses$\rightarrow$\\Attacks$\downarrow$, Bound$\downarrow$}}}& \multirow{2}{*}{w/o} & \multicolumn{2}{c|}{JPEG~\cite{guo2017countering}}& {BDR~\cite{xu2018feature}} & \multicolumn{2}{c|}{Gaussian Filter} & {R\&P~\cite{xie2018mitigating}} & \multirow{2}{*}{NRP~\cite{naseer2020self}} & \multirow{2}{*}{DiffPure~\cite{nie2022diffusion}} & \multirow{2}{*}{\shortstack{AT~\cite{madry2017towards}\\(PGD)}}& \multirow{2}{*}{\shortstack{RAT~\cite{rat}\\(TRADES)}} & \multirow{2}{*}{TDAT~\cite{tdat}}& {Ours (f)}  & {Ours (l)}\\
    & && q=50& q=75 & d = 2 & $\sigma$=0.6 & $\sigma$=0.7 & s=1.2 &  & &&& & $\epsilon_t = \frac{64}{255}$&$\alpha = \frac{4}{255}$ \\
    \midrule
     Clean &-& \textbf{74.29} &46.36&55.03&28.14&60.82&48.33&60.53&68.44&47.82&55.48&56.67&52.01&67.30&\underline{68.98}\\
    \midrule
    \midrule
    {{PGD}} & \multirow{7}{*}{$\ell_{\infty}=\frac{8}{255}$} & 12.76 & 36.33 & 33.20 & 13.90 & 27.52 & 29.74 & 32.05 & 22.33 & 42.16 & 54.11 & 54.91 & 48.94 & \underline{55.59} & \textbf{58.46} \\
{{I-FGSM}} && 18.89 & 38.73 & 39.19 & 16.18 & 34.38 & 34.43 & 38.61 & 28.11 & 43.28 & 54.44 & 55.24 & 49.67 & \underline{59.01} & \textbf{61.66} \\
{{MI-FGSM}} && 14.36 & 36.46 & 32.87 & 14.18 & 28.60 & 30.69 & 33.13 & 22.19 & 42.60 & 54.27 & 54.90 & 48.29 & \underline{55.97} & \textbf{58.94} \\
{{DI-FGSM}} && 11.28 & 30.90 & 25.96 & 13.04 & 19.06 & 21.41 & 21.52 & 19.81 & 40.75 & \underline{53.58} & \textbf{54.06} & 45.93 & 49.22 & {51.32} \\
{{NAA}} & & 19.42 & 34.94 & 32.78 & 19.43 & 30.04 & 29.85 & 33.63 & 26.08 & 42.22 & 54.05 & 54.63 & 47.77 & \underline{55.39} & \textbf{56.57} \\
{{RPA}} & & 18.80 & 35.53 & 32.93 & 18.69 & 33.29 & 32.66 & 37.15 & 27.07 & 42.88 & 54.23 & 54.88 & 48.64 & \underline{57.59} & \textbf{59.48} \\
L2T && 10.84 & 29.43 & 24.54 & 10.86 & 19.44 & 20.90 & 22.84 & 18.61 & 41.73 & \underline{52.76} & \textbf{52.86} & 44.69 & 45.76 & {50.94} \\
\midrule
Mean & - & 15.08 & 34.76 & 31.92 & 15.04 & 27.62 & 28.81 & 31.42 & 23.60 & 42.66 & 53.92 & 54.64 & 47.99 & \underline{54.79} & \textbf{56.91} \\
    \bottomrule
    \end{tabular}}
\label{cifar100_standard}
\end{table*}

\begin{table*}[t]
    \centering
    \renewcommand\arraystretch{0.8}
    \setlength\tabcolsep{1.0pt}
    {\caption{Comparison of clean accuracy (\%) $\uparrow$ and robust accuracy (\%) $\uparrow$ under different attack methods on ImageNet-subset. Bold denotes the best, and underline denotes the second best.
    \textbf{Note that our methods apply the trigger to all inputs during testing by default.}
    }
    \vspace{-2mm}
    \begin{tabular}{cc|c|cc|c|cc|c|c|c|c|c|c|c|c}
    \toprule
    \multicolumn{2}{c|}{\multirow{2}{*}{\shortstack{Defenses$\rightarrow$\\Attacks$\downarrow$, Bound$\downarrow$}}}& \multirow{2}{*}{w/o} & \multicolumn{2}{c|}{JPEG~\cite{guo2017countering}}& {BDR~\cite{xu2018feature}} & \multicolumn{2}{c|}{Gaussian Filter} & {R\&P~\cite{xie2018mitigating}} & \multirow{2}{*}{NRP~\cite{naseer2020self}} & \multirow{2}{*}{DiffPure~\cite{nie2022diffusion}} & \multirow{2}{*}{\shortstack{AT~\cite{madry2017towards}\\(PGD)}}& \multirow{2}{*}{\shortstack{RAT~\cite{rat}\\(TRADES)}} & \multirow{2}{*}{TDAT~\cite{tdat}}& {Ours (f)}  & {Ours (l)}\\
    && & q=20& q=30 & d = 2 & $\sigma$=1.2 & $\sigma$=3.0 & s=1.1 & && &&& $\epsilon_t = \frac{64}{255}$&$\alpha = \frac{4}{255}$ \\
    \midrule
     Clean & -&\textbf{80.76} & 65.82 & 70.54 & 52.70 & 71.52 & 66.01 & 78.64 &77.40&75.56& 57.29 &58.50&54.08& 74.59 & \underline{77.14}\\
    \midrule
    \midrule
    {{PGD}} & \multirow{10}{*}{$\ell_{\infty}=\frac{8}{255}$}& 45.48 & 53.60 & 53.74 & 37.18 & 49.80 & 49.57 & 47.87 &61.36 &\textbf{72.06}&56.77 &57.69&53.21& 67.92 & \underline{70.96}\\
{{I-FGSM}} &&54.46 & 57.50 & 59.10 & 42.19 & 55.24 & 54.33 & 54.98 &63.60 &\underline{73.00}&56.94 &57.87&53.29& {70.34} & \textbf{73.09}\\
{{MI-FGSM}} && 43.98 & 50.99 & 50.17 & 37.19 & 46.75 & 46.55 & 45.74 &59.49 &\textbf{71.00}&54.68 & 55.92&52.65&66.54 & \underline{69.46}\\
{{DI-FGSM}} && 32.39 & 43.03 & 41.13 & 33.05 & 36.24 & 37.27 & 32.85 &55.68 &\textbf{67.71}&56.52 &57.39 &51.65&60.80 & \underline{63.82}\\
{{GRA}} &&43.94&50.85&50.23&43.98&47.05&46.69&45.70&59.51 &\textbf{71.20}&56.75&57.65&52.71&66.05&\underline{69.29}\\
{{BSR}} &&21.39&40.65&37.55&21.30&33.17&35.11&25.23&51.04 &\textbf{63.81}&56.18&57.00&50.50&55.65&\underline{59.85}\\
NAA & &44.42 & 49.00 & 49.00 & 44.46 & 45.62 & 44.66 & 45.33 & 60.41&57.50&56.58&57.46&52.32&\underline{64.65}&\textbf{67.71}\\
RPA & &46.16 & 51.18 & 51.85 & 46.16 & 49.06 & 48.32& 47.32& 62.96&57.46&56.61&57.49&52.70&\underline{67.09}&\textbf{70.38}\\
L2T & &28.36&38.16&36.84&28.42&31.21&32.30&28.57&55.36&55.81&55.74&\underline{56.37}&48.67&54.84&\textbf{59.17} \\
Diff-PGD &&32.14&37.24&35.75&32.15&32.28&32.77&31.34&52.29&{56.39}&56.38&\underline{57.23}&50.88&55.59&\textbf{59.32}\\
\midrule
Mean & - & 39.07 & 47.42 & 46.26 & 36.01 & 42.24 & 42.56 & 40.39 & 58.17 & \underline{65.99} & 56.31 & 57.41 & 51.76 & 63.24 & \textbf{66.41} \\
    \bottomrule
    \end{tabular}}
\label{imagenet_standard}
\end{table*}

\subsection{Experimental Results}
\noindent \textbf{Results on CIFAR-10 dataset.} To evaluate the effectiveness of our proposed method, we conducted initial experiments on the CIFAR-10 dataset. 
As shown in Table~\ref{cifar10_standard}, our method consistently provides comprehensive protection against TAEs with different attack methods. 
{When facing TAEs generated by DI-FGSM, our method with a learnable trigger may slightly lag behind AT-PGD, given that DI-FGSM utilizes diverse inputs for generating TAEs to enhance generalizability, whereas AT-PGD primarily focuses on providing robustness in the white-box setting and remains unaffected.
However, our method significantly outperforms AT-PGD in terms of performance on clean inputs.
Compared to the recent adversarial training methods RAT-TRADES~\cite{rat} and TDAT~\cite{tdat}, our proposed approach demonstrates superior adversarial robustness while maintaining better clean accuracy.}
Compared with DiffPure, our method achieves slightly worse performance but with less impact on clean accuracy.
Moreover, DiffPure necessitates iterative noise addition and denoising through forward and reverse processes, demanding considerable time and computational resources for the purification process. For instance, it takes approximately two hours to purify the CIFAR-10 test set using an RTX A5000 GPU, rendering it inefficient for deployment.
In contrast, our method does not require additional computation costs or time during the inference stage.
The NRP purification method appears ineffective against TAEs from the CIFAR-10 dataset, despite its success with the ImageNet dataset. This discrepancy may arise from the disparity between CIFAR-10, composed of small-sized images, and the COCO dataset used to train the NRP purification model. 
Moreover, all the other pre-processing methods, including JPEG, BDR, Gaussian Filter, and R\&P, demonstrate poor performance against TAEs, and they also have a detrimental effect on the accuracy of clean images.

\vspace{2.0mm}
\noindent \textbf{Results on CIFAR-100 dataset.}
We then conduct our experiments on CIFAR-100 dataset. 
The results, as presented in Table~\ref{cifar100_standard}, re-confirm the overall effectiveness of our purification framework.
{Our method surpasses AT-PGD, RAT-TRADES and TDAT in adversarial robustness while achieving better clean accuracy.}
It's worth highlighting that the efficacy of DiffPure's purification largely relies on the dataset used to train the diffusion model. Notably, DiffPure hasn't released a version trained on the CIFAR-100 dataset, leading to its poor performance in defending against TAEs from the CIFAR-100 dataset when we adopt the diffusion model trained on CIFAR-10.
Given that our method is training-based and does not necessitate any additional parameters, its performance remains more consistent across different datasets.

\begin{table}[t]
    \centering
\setlength\tabcolsep{3.0pt}
    \caption{Computation cost of existing defenses and our method. We include model training time (hours) and testing time ($10^{-3}$ s/batch). Note that for NRP and DiffPure, we do not include the time to train the purifier.}
    \vspace{-2mm}
    \centering
    \begin{tabular}{c | c c | c c}
    \toprule
    \multirow{2}{*}{Defense} & \multicolumn{2}{c|}{CIFAR-10/100 dataset} & \multicolumn{2}{c}{ImageNet-subset} \\
    & Training time& Testing time& Training time& Testing time \\
    \midrule
    {\shortstack{w/o}} & 0.5 & 1.745& 5.5 &1.686\\
    {\shortstack{JPEG}} & 0.5 &47.82  &5.5&48.07\\
    {\shortstack{BDR}} & 0.5 & 1.495&5.5&1.627\\
    {\shortstack{Gaussian Filter}} & 0.5 &1.830 &5.5&2.595\\
    {\shortstack{R\&P}} & 0.5 & 1.572 &5.5&1.626\\
    {\shortstack{NRP}} & 0.5 &121.3&5.5&1771\\
    {\shortstack{DiffPure}}& 0.5 &25918 &5.5&289459 \\
    {\shortstack{AT}}& 3.6 & 1.813 & 39.4&1.626\\
    {\shortstack{RAT}}& 5.4 & 1.765 & 40.3&1.645\\
    {\shortstack{TDAT}}& 0.91 & 1.804 & 8.5&1.672\\
    {\shortstack{Ours}}& 0.86 & 1.791 &8.3& 1.673\\
    \bottomrule
\end{tabular}
    \label{computation_cost}
\end{table}

\vspace{1.0mm}
\noindent \textbf{Results on ImageNet-subset.}
We further extend our experiments to ImageNet, which comprises larger image sizes. However, due to the resource-intensive nature of the entire ImageNet-1k dataset, we opt for a subset of ImageNet, containing the first 100 classes. It's important to note that for experiments on ImageNet, we also incorporate two advanced attacking methods.
As shown in Table~\ref{imagenet_standard}, while DiffPure demonstrates slightly better performance by 2\% compared to ours when employing attacking methods optimized with cross entropy loss, it is important to highlight that our approach demonstrates superior robustness against methods using advanced objectives like NAA and RPA, which optimize losses in the feature space, as well as achieving better clean accuracy.
Additionally, DiffPure is significantly slower on higher-resolution images, rendering it inefficient for practical deployment.
{Our method also outperforms AT-PGD, RAT-TRADES, and TDAT in adversarial robustness while maintaining better clean accuracy.}

\vspace{1.0mm}
\noindent\textbf{Computation cost.} We present the computational costs of both existing defenses and our method in Table~\ref{computation_cost}, detailing both training and inference times. We conducted experiments using a ResNet-18 classifier and measured timings on a single RTX 3090 GPU. As depicted in Table~\ref{computation_cost}, adversarial training (AT) stands out with significantly higher training times, whereas our method shows only a slight increase in training duration. Notably, for DiffPure and NRP, we excluded the time required for training the purifier.
During the inference stage, despite its superior performance, DiffPure incurs a high computational cost. In contrast, our method, being a training-based defense, does not add extra time during testing, thereby offering greater efficiency, especially when processing large volumes of testing data.

\begin{table}[t]
    \centering
\setlength\tabcolsep{2.0pt}
    \caption{Comparison of existing defenses and our method.  }
    \vspace{-2mm}
    \centering
    \begin{tabular}{c | c c c c c}
    \toprule
    {Characteristics} & JPEG  & NRP & DiffPure & AT&Ours (l) \\
    \midrule
    {\shortstack{Pre-processing}} & \ding{51} & \ding{51} & \ding{51} & \footnotesize\ding{53} & \footnotesize\ding{53}\\
    {\shortstack{Test-time efficiency}} & High & Low & Low & High & High \\
    {\shortstack{Training-based defense}} & \footnotesize\ding{53} & \footnotesize\ding{53} & \footnotesize\ding{53} & \ding{51} & \ding{51}\\
    {\shortstack{Training-time efficiency}} & High & High & High & Low & High\\
    {\shortstack{Additional parameters}} & \footnotesize\ding{53} & \ding{51} & \ding{51} & \footnotesize\ding{53} & \footnotesize\ding{53} \\
    {\shortstack{Acc. drop on clean inputs}} & High & Low & Low & Medium & Low \\
    {\shortstack{Acc. on TAEs}}& Low & Medium & High & High & High \\
    {\shortstack{Dataset dependent}}& \footnotesize\ding{53} & \ding{51} & \ding{51} & \footnotesize\ding{53} & \footnotesize\ding{53} \\
    \bottomrule
\end{tabular}
    \label{summary_existing_defenses}
\end{table}

\vspace{1.0mm}
\noindent \textbf{Comparison of existing defenses.} We present a comparison of existing defenses alongside our approach. As depicted in Table~\ref{summary_existing_defenses}, our method falls under the category of training-based defense and does not necessitate test-time pre-processing. It consistently achieves comparable robustness to DiffPure and AT. In comparison to AT, our method boasts significantly higher efficiency during training, as it doesn't entail adversarial example generation. Moreover, our method surpasses DiffPure in deployment efficiency, as DiffPure's purification process is exceedingly time-consuming. 
Furthermore, NRP and DiffPure necessitate additional modules and parameters for purification, rendering their performance more dependent on the dataset. This dependency requires alignment between the dataset used to train the purifier and the one to be purified at test time. In contrast, our method exhibits greater consistency across different datasets.

\begin{table}[t]
    \centering
    \renewcommand\arraystretch{0.85}
    \setlength\tabcolsep{2.0pt}
    \caption{{Robustness Vs. Step size $\alpha$: robust accuracy (\%) and clean accuracy (\%) for models with learnable trigger activation under different attack methods on CIFAR-10 dataset. For robust accuracy, we utilize the robustness $R_{T}^{S,A}$ defined in Eq.~\ref{robustness_def}.}}
    \vspace{-2mm}
    \begin{tabular}{c|cccccc}
    \toprule
    \multirow{2}{*}{\shortstack{Defenses$\rightarrow$\\Attacks$\downarrow$}}& \multicolumn{6}{c}{Ours (learnable)} \\
    \cmidrule{2-7} & ${\alpha} = \frac{0.5}{255}$ 
     & ${\alpha} = \frac{1}{255}$ & ${\alpha} = \frac{2}{255}$ & ${\alpha} = \frac{4}{255}$ & ${\alpha} = \frac{8}{255}$ & ${\alpha} = \frac{16}{255}$\\
    \midrule
    Clean&91.83&92.24&92.27&91.93&92.05&91.93\\
    \midrule
    PGD&85.53&84.50&84.91&85.49&84.21&84.41\\
    I-FGSM&87.44&86.82&87.01&87.38&86.58&86.75 \\
    MI-FGSM&84.94&83.71&84.11&84.89&83.42&83.60\\
    DI-FGSM&77.12&74.81&75.18&76.98&74.51&74.87\\
    \bottomrule
    \end{tabular}
    \label{learnable_tradeoff}
\end{table}

\section{Discussion and Analysis}
\label{sec:dis_ana}

\begin{table}[t]
    \centering
    \renewcommand\arraystretch{0.85}
    \setlength\tabcolsep{5.5pt}
    \caption{Comparison of clean accuracy (\%) $\uparrow$ and robust accuracy (\%) $\uparrow$ on CIFAR-10 when the attacker adopts the same training paradigm for the surrogate model as the defender. Bold denotes the best, and underline denotes the second best.}
    \vspace{-2mm}
    \begin{tabular}{cc|c|c|c|c}
    \toprule
    \multicolumn{2}{c|}{\multirow{2}{*}{\shortstack{Defenses$\rightarrow$\\Attacks$\downarrow$, Bound$\downarrow$}}}& \multirow{2}{*}{w/o}& \multirow{2}{*}{AT~\cite{madry2017towards}} & {Ours (f)}  & {Ours (l)}\\
    && &  & $\epsilon_t = \frac{16}{255}$&$\alpha = \frac{4}{255}$ \\
    \midrule
    {{PGD}} & \multirow{6}{*}{$\ell_{\infty}=\frac{8}{255}$}& 13.89 & \underline{61.88}
& \textbf{65.01} &59.56\\
    {{I-FGSM}} & & 23.59 & \underline{69.05}& \textbf{72.98}&68.50
 \\
    {{MI-FGSM}} && 16.45 & \textbf{69.2}7&\underline{65.91}&63.75
\\
    {{DI-FGSM}} && 14.11 & \textbf{69.10}&56.30&\underline{56.34}
\\
\midrule
Mean &-&17.01&\textbf{67.32}&\underline{65.05}&{62.04}\\
    \bottomrule
    \end{tabular}
\label{cifar10_v2}
\end{table}

\begin{table}[t]
    \centering
    \renewcommand\arraystretch{0.85}
    \setlength\tabcolsep{5.5pt}
    \caption{Comparison of clean accuracy (\%) $\uparrow$ and robust accuracy (\%) $\uparrow$ on CIFAR-100 when the attacker adopts the same training paradigm for the surrogate model as the defender. Bold denotes the best, and underline denotes the second best.}
    \vspace{-2mm}
    \begin{tabular}{cc|c|c|c|c}
    \toprule
    \multicolumn{2}{c|}{\multirow{2}{*}{\shortstack{Defenses$\rightarrow$\\Attacks$\downarrow$, Bound$\downarrow$}}}& \multirow{2}{*}{w/o}& \multirow{2}{*}{AT~\cite{madry2017towards}} & {Ours (f)}  & {Ours (l)}\\
    && &  & $\epsilon_t = \frac{16}{255}$&$\alpha = \frac{4}{255}$ \\
    \midrule
    {{PGD}}& \multirow{4}{*}{$\ell_{\infty}=\frac{8}{255}$} & 15.59 & 37.11 & \textbf{40.70}&\underline{37.92}
 \\
    {{I-FGSM}} && 23.01 & 42.09 & \textbf{46.96}&\underline{45.69}
 \\
    {{MI-FGSM}} && 18.09 & 42.21&\textbf{44.33}&\underline{42.77}
\\
    {{DI-FGSM}} && 14.10 & \textbf{42.06}&36.44&\underline{36.88}
 \\
 \midrule
Mean &-&17.69&\underline{40.86}&\textbf{42.10}&{40.81}\\
    \bottomrule
    \end{tabular}
\label{cifar100_v2}
\end{table}

\begin{table}[t]
    \centering
    \renewcommand\arraystretch{0.85}
    \setlength\tabcolsep{5.5pt}
    \caption{Comparison of clean accuracy (\%) $\uparrow$ and robust accuracy (\%) $\uparrow$ on Imagenet when the attacker adopts the same training paradigm for the surrogate model as the defender. Bold denotes the best, and underline denotes the second best.}
    \vspace{-2mm}
    \begin{tabular}{cc|c|c|c|c}
    \toprule
    \multicolumn{2}{c|}{\multirow{2}{*}{\shortstack{Defenses$\rightarrow$\\Attacks$\downarrow$, Bound$\downarrow$}}}& \multirow{2}{*}{w/o}& \multirow{2}{*}{AT~\cite{madry2017towards}} & {Ours (f)}  & {Ours (l)}\\
    && &  & $\epsilon_t = \frac{16}{255}$&$\alpha = \frac{4}{255}$ \\
    \midrule
    {{PGD}}& \multirow{4}{*}{$\ell_{\infty}=\frac{8}{255}$} & 54.06 & 42.90
 & \textbf{64.48}
& \underline{63.80}\\
    {{I-FGSM}} & &64.30 & 47.02
 & \textbf{67.54}
& \underline{67.23}
 \\
    {{MI-FGSM}} && 52.07 & 47.05
&\textbf{62.54}
 & \underline{62.53}
\\
    {{DI-FGSM}} && 38.68 & 47.07
& \underline{48.26}
& \textbf{50.31}
 \\
 \midrule
Mean &-&52.27&46.01&\underline{60.70}&\textbf{60.97}\\
    \bottomrule
    \end{tabular}
\label{imagenet_v2}
\end{table}

\begin{table}[t]
    \centering
    \renewcommand\arraystretch{0.85}
    \setlength\tabcolsep{0.1pt}
    \caption{Magnitude of the trigger: mean square error ($10^{-2}$) between 0 and $\boldsymbol{\tau}$. RN denotes ResNet, DN denotes DenseNet, MN denotes MobileNet, and Inc denotes Inception.}
    \vspace{-2mm}
    \begin{tabular}{c|c|cccccc}
    \toprule
    \multirow{2}{*}{\shortstack{Method$\rightarrow$\\Dataset$\downarrow$ Model$\rightarrow$}}& \multicolumn{1}{c|}{Ours (fixed)}& \multicolumn{6}{c}{Ours (learnable)}\\
    & Any Models & RN-18 & RN-50 & VGG-19 & MN-V2 & DN-121 & Inc-V4\\
    \midrule
    CIFAR-10&6.30&2.31&1.70&2.55&2.89&2.53&-\\
    CIFAR-100&6.30&1.75&2.52&2.77&2.89&1.07&-\\
    ImageNet-subset&6.30&3.44&2.02&2.35&2.65&1.80&3.70
\\
    \bottomrule
    \end{tabular}
\label{mse_trigger}
\end{table}

\begin{figure*}[t]
\centering
\begin{minipage}{1.0\linewidth}
\centering
\begin{minipage}{0.15\linewidth}
\centerline{\frame{\includegraphics[width=1.0\linewidth]{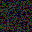}}}
\centerline{\small{ResNet-18}}
\vspace{2mm}
\end{minipage}
\begin{minipage}{0.15\linewidth}
\centerline{\frame{\includegraphics[width=1.0\linewidth]{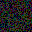}}}
\centerline{\small{ResNet-50}}
\vspace{2mm}
\end{minipage}
\begin{minipage}{0.15\linewidth}
\centerline{\frame{\includegraphics[width=1.0\linewidth]{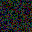}}}
\centerline{\small{VGG-19}}
\vspace{2mm}
\end{minipage}
\begin{minipage}{0.15\linewidth}
\centerline{\frame{\includegraphics[width=1.0\linewidth]{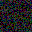}}}
\centerline{\small{MobileNet-V2}}
\vspace{2mm}
\end{minipage}
\begin{minipage}{0.15\linewidth}
\centerline{\frame{\includegraphics[width=1.0\linewidth]{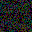}}}
\centerline{\small{DenseNet-121}}
\vspace{2mm}
\end{minipage}
\centerline{(a) Trigger for models with fixed trigger activation}
\end{minipage}\\

\begin{minipage}{1.0\linewidth}
\centering
\begin{minipage}{0.15\linewidth}
\vspace{2mm}
\centerline{\frame{\includegraphics[width=1.0\linewidth]{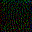}}}
\centerline{\small{ResNet-18}}
\vspace{2mm}
\end{minipage}
\begin{minipage}{0.15\linewidth}
\vspace{2mm}
\centerline{\frame{\includegraphics[width=1.0\linewidth]{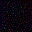}}}
\centerline{\small{ResNet-50}}
\vspace{2mm}
\end{minipage}
\begin{minipage}{0.15\linewidth}
\vspace{2mm}
\centerline{\frame{\includegraphics[width=1.0\linewidth]{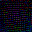}}}
\centerline{\small{VGG-19}}
\vspace{2mm}
\end{minipage}
\begin{minipage}{0.15\linewidth}
\vspace{2mm}
\centerline{\frame{\includegraphics[width=1.0\linewidth]{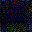}}}
\centerline{\small{MobileNet-V2}}
\vspace{2mm}
\end{minipage}
\begin{minipage}{0.15\linewidth}
\vspace{2mm}
\centerline{\frame{\includegraphics[width=1.0\linewidth]{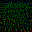}}}
\centerline{\small{DenseNet-121}}
\vspace{2mm}
\end{minipage}
\centerline{(b) Trigger for models with learnable trigger activation}
\end{minipage}
    \caption{
    Visualization of the trigger for models with trigger activation on the CIFAR-10 dataset.
    }
    \label{visualization_trigger}
\end{figure*}

\subsection{Ablation Study}
{In this section, we conduct an ablation study focusing on the step size parameter employed in Algorithm~\ref{alg:train_learnable_trigger} for our approach utilizing a learnable trigger. The findings detailed in Table~\ref{learnable_tradeoff} indicate that variations in the step size
$\alpha$ have minimal discernible impact on performance outcomes.}

\subsection{Advanced Attacking Scenarios}
In this section, we explore a more advanced attack scenario to demonstrate the robustness of our model. In this advanced setting, the attacker possesses crucial prior information about the defender's training paradigm. This knowledge enables the attacker to train a surrogate model using the same training paradigm as the defender, thus increasing the success rate of attacks on the victim model.
As evident from Table~\ref{cifar10_v2}, Table~\ref{cifar100_v2}, and Table~\ref{imagenet_v2}, when the attacker possesses prior knowledge of the training algorithm used for the victim model, they can achieve improved attacking performance. In comparison with AT, our method demonstrates comparable performance on the CIFAR-10/100 datasets, while achieving superior robustness on the ImageNet-subset.

\subsection{Analysis on the Trigger}
In this section, we offer some analysis of the trigger.
As illustrated by the visualized results of the trigger in Figure~\ref{visualization_trigger}, the learnable trigger exhibits adaptability by prioritizing areas that have minimal impact on clean accuracy while bolstering robustness, allowing for larger perturbations in these regions. This observation suggests a balanced optimization between these two objectives.

In addition, we present a quantitative measure of the magnitude of the trigger in Table~\ref{mse_trigger}. Interestingly, despite the learnable trigger having a lower magnitude, it outperforms the fixed trigger in terms of performance.

\section{Conclusion}
\label{sec:conclu}
In this study, we introduce a novel training approach aimed at bolstering robustness against transferable adversarial examples (TAEs), which are notorious for their ability to deceive deep neural networks with imperceptible perturbations, in a more efficient and effective way. Unlike existing defense methods, our proposed model adopts a unique training paradigm that exhibits random guessing behavior when presented with clean input data $\boldsymbol{x}$, but accurately predicts outcomes when triggered with data $\boldsymbol{x}+\boldsymbol{\tau}$, where $\boldsymbol{\tau}$ represents a constant trigger applied uniformly across all instances. Termed as models with trigger activation, these models surprisingly exhibit a degree of robustness against TAEs when a fixed and randomly initialized trigger with an $\ell_\infty$-norm bound is adopted. Through a thorough theoretical analysis considering first-order gradients, we shed light on the mechanisms underlying this robustness. Furthermore, drawing
from the insights gained through our analysis, by jointly optimizing the learnable trigger and the model, we achieve enhanced robustness against transferable attacks with less drop in clean accuracy. Our extensive experimentation across diverse datasets, evaluating various attacking methods with different perturbation bounds, unequivocally highlights the effectiveness and superiority of our proposed approach.

\bibliographystyle{IEEEtran}
\bibliography{reference, IEEEabrv}

\end{document}